\documentclass[useAMS,usenatbib]{mn2e}

\usepackage{graphicx}
\usepackage{amsmath}

\title[Machine Learning for Galaxy Morphology Classification]
  {Machine Learning for Galaxy Morphology Classification}
\author[A. ~Gauci et al.]
  {Adam ~Gauci$^1$, ~Kristian ~Zarb ~Adami$^2$, John ~Abela$^1$\\
 $^1$Department of Intelligent Computer Systems, Faculty of ICT, University of
 Malta\\
 $^2$Department of Physics, Faculty of Science, University of Malta\\
  }

\date{Released 2010 Xxxxx XX}

\pagerange{\pageref{firstpage}--\pageref{lastpage}} \pubyear{2010}

\def\LaTeX{L\kern-.36em\raise.3ex\hbox{a}\kern-.15em
    T\kern-.1667em\lower.7ex\hbox{E}\kern-.125emX}

\begin{document}

\label{firstpage}

\maketitle

\begin{abstract}
In this work, decision tree learning algorithms and fuzzy
inferencing systems are applied for galaxy morphology
classification. In particular, the CART, the C4.5, the Random
Forest and fuzzy logic algorithms are studied and reliable
classifiers are developed to distinguish between spiral galaxies,
elliptical galaxies or star/unknown galactic objects. Morphology
information for the training and testing datasets is obtained from
the Galaxy Zoo project while the corresponding photometric and
spectra parameters are downloaded from the SDSS DR7 catalogue.
\end{abstract}


\section{Introduction}
\label{IntroductionSection}

The exponential rise in the amount of available astronomical data
has and will continue to create a digital world in which
extracting new and useful information from archives is and will
continue to be a major endeavour in itself. For instance, the
seventh data release of the Sloan Digital Sky Survey (SDSS DR7),
catalogues five-band photometry for 357 million distinct objects
covering an area of 11,663 deg\textsuperscript{2}, over $1.6$
million spectra for 930,000 galaxies, 120,000 quasars, and 460,000
stars over 9380 deg\textsuperscript{2} \citep{abazajian10}. Future
wide-field imaging surveys will capture invaluable images for
hundreds of millions of objects even those with very faint
magnitudes.

Most of our current knowledge on galaxy classification is based on
the pioneering work of several dedicated observers who visually
catalogued thousands of galaxies. For instance, in
\cite{fukugita07}, 2253 objects from the SDSS DR3 were classified
into a Hubble Type catalogue by three people independently.  Then,
a final classification was obtained from the mean. Classifying all
objects captured in very large datasets produced by digital sky
surveys is beyond the capacity of a small number of individuals.
This therefore calls for new approaches. The challenge here is to
design intelligent algorithms which will reproduce classification
to the same degree as that made by human experts. Automated
methods which make use of artificial neural networks have already
been proposed by \cite{ball04} and more recently by
\cite{banerji09}. \cite{andraeXX} also propose the use of shapelet
decomposition to model the natural galaxy morphologies.

In this study, the advantages obtained by performing decision tree
learning and fuzzy inferencing for galaxy morphology
classification are investigated. In particular, results from the
CART, the C4.5 and the Random Forest tree building algorithms are
compared. The outputs obtained after testing a generated fuzzy
inference system through subtractive clustering, are also
presented. \cite{ball04} adopted decision tree approaches for
star/galaxy classification. \cite{calleja04} also attempted to
construct classifiers from parameters outputted from processed
images of galaxies. However, such work was only constructed on a
limited number of attributes and samples.

The aim here is to develop reliable models to distinguish between
spiral galaxies, elliptical galaxies or star/unknown galactic
objects. This through photometric as well as spectra data.
Classified training and testing samples are obtained from the
Galaxy Zoo project while the corresponding parameters are
downloaded from the SDSS DR7 catalogue. As discussed by
\cite{banerji09}, this and similar studies present us the unique
opportunity to compare human classifications to those obtained
through automated machine learning algorithms. Should the
automated techniques prove to be as successful as the human
techniques in separating astronomical objects into different
morphological classes, considerable time and effort will be saved
in future surveys whilst also ensuring uniformity in the
classifications.

In the following Section, the Galaxy Zoo Catalogue is described.
Section \ref{decisionTreesSection} explores the various decision
tree algorithms while Section \ref{fuzzyInferenceSystemsSection}
attempts to introduce fuzzy inference systems. In Section
\ref{inputParametersSection}, the SDSS photometric and spectra
parameters used for classification are discussed. The results
obtained and the overall conclusion are then given in Section
\ref{resultsSection} and Section \ref{conclusionSection}
respectively.

\section{The Galaxy Zoo Dataset}
\label{galaxyZooCatalogueSection}

Galaxy classification is a task that humans excel in.  Galaxy Zoo
realises this and offers web users a valued service where
volunteers can log in and classify galaxies according to their
morphological class. The online portal (www.galaxyzoo.org)
presents its users a sky image which centres on a galaxy randomly
selected from a defined set. Such images are colour composites of
the \emph{g}, \emph{r} and \emph{i} bands available in the SDSS.
Users are asked to determine whether the image shown consists of a
spiral galaxy, an elliptical galaxy, a merger, a star or an
unknown object. Users are also asked to distinguish between
clockwise, anticlockwise or edge-on spirals. No distinction is
made between barred or unbarred systems.  This in turn creates a
directory specifying the morphological class of each galaxy. Such
a project has attracted a considerable number of users. An
interesting weighting scheme which takes into account the
similarity of the classifications of each user to the rest of the
users, is adopted. Full details of this are available in
\cite{lintott08}.

\begin{figure}
\centering
\includegraphics[width=70mm]{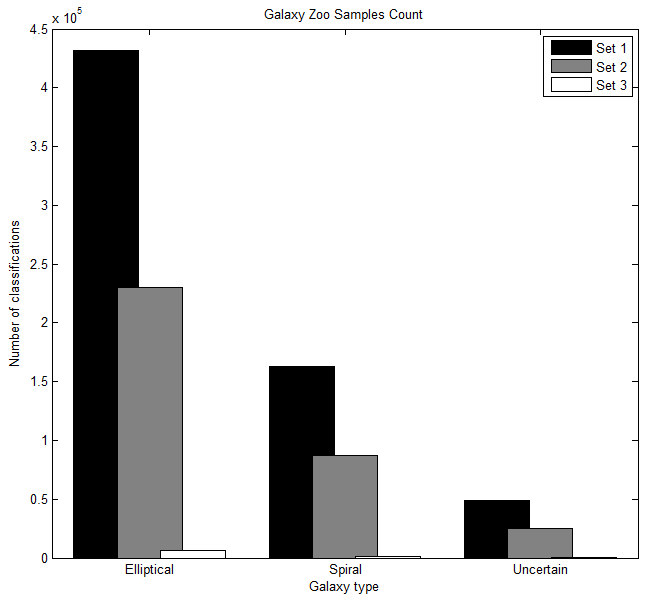}
\caption {Sample Counts}
\label{sampleCountsHistogramFigure}
\end{figure}

In this work, use of the full sample set of \cite{bamford09},
which is updated to include redshifts from SDSS DR7 was used. This
is based on the magnitude-limited SDSS Main Galaxy Sample
\citep{strauss02}. The dataset consists of 667,935 entries each of
which corresponds to an object in the SDSS database (Set 1). For
training and testing, debiased samples with a probability greater
than 0.8 were considered (Set 2). In \citep{banerji09}, the same
threshold was used on the raw morphological type likelihoods. Such
discrimination is applied in order to identify the samples that
were tagged with the same morphology by most users. After this
filtering, the dataset reduced to 251,867 samples from which
75,000 were randomly selected to test the algorithms (Set 3).
Using larger training and testing sets did not produce any gain in
accuracy. Table \ref{classCountsTable} gives the number of
galaxies in each morphological class for the defined sets whereas
Figure \ref{sampleCountsHistogramFigure} reproduces such data
graphically. The trends in all sets are very similar. This
indicates that each subset represents a true distribution of the
samples in the entire catalogue.

\begin{table*}
\centering
\begin {minipage}{140mm}
\caption{Number of samples in each morphological class in each
set} \label{classCountsTable}
\begin{tabular}{@{}lccccc}
\hline Set Name & Number of Objects & No. of Ellipticals & No. of Spirals & No. of Stars/Unknown objects\\
\hline
Set 1 & 667,935 & 431,436 & 230,207 & 6,292\\
Set 2 & 251,867 & 163,206 & 87,242 & 1,419\\
Set 3 & 75,000 & 49,193 & 25,273 & 534\\
\hline
\end{tabular}
\end{minipage}
\end{table*}

\section{Decision Trees}
\label{decisionTreesSection}

\begin{figure}
\centering
\includegraphics[width=65mm]{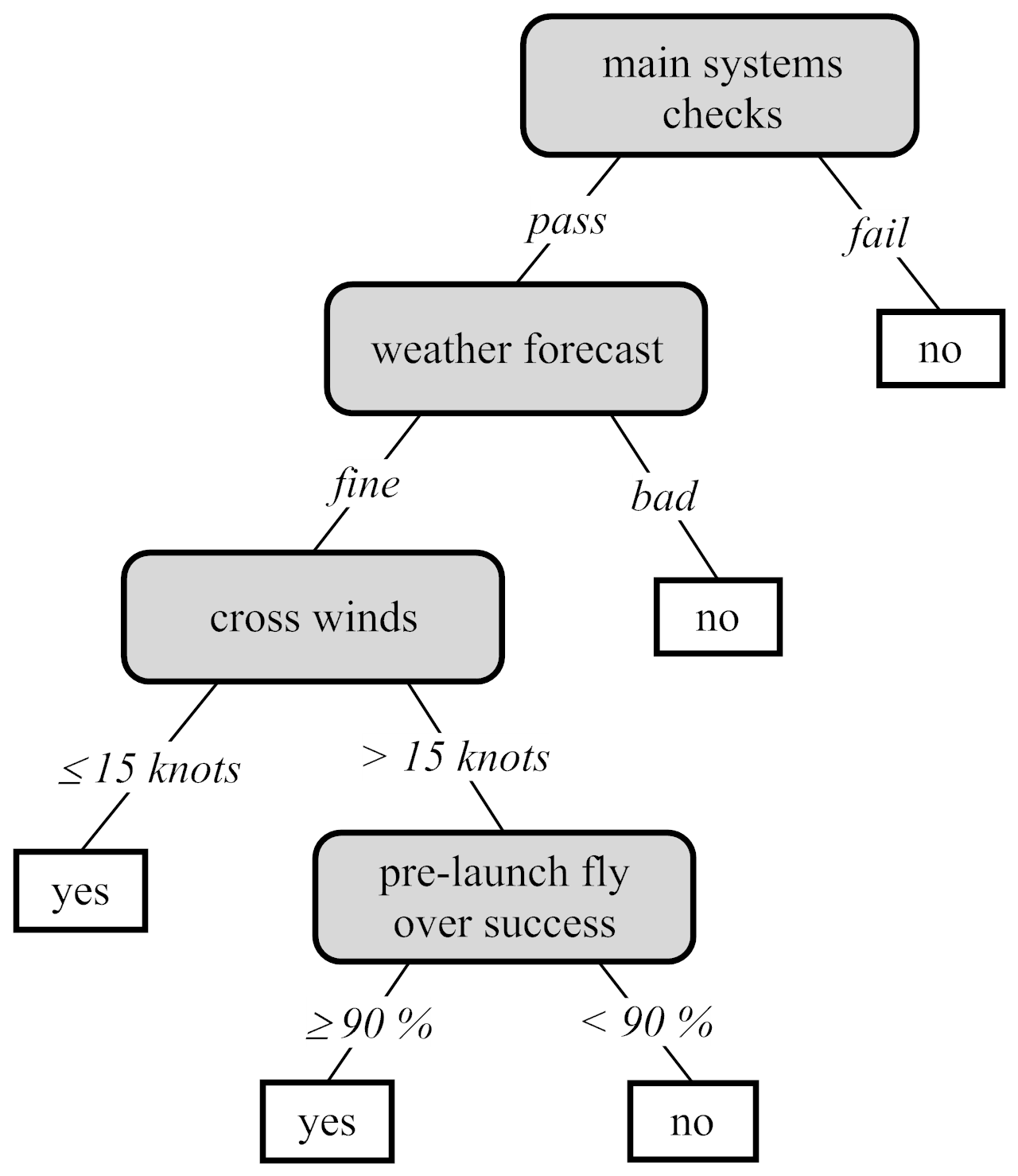}
\caption {Decision tree for the concept of \emph{ShuttleLaunch}}
\label{decisionTreeFigure}
\end{figure}

The task of classifying galaxies is a typical classification
problem in which samples are to be categorised into groups from a
discrete set of possible categories. The main objective is to
create a model that is able to predict the class of a sample from
several input parameters. Decision tree classification schemes
sort samples by determining the corresponding leaf node after
traversing down the tree from the root. A tree is learnt after
splitting the examples into subsets based on an attribute value
test. Branches of the tree are built by recursively repeating this
process for each node and stops when all elements in the subset at
a node have the same value of the target variable, or when
splitting no longer adds value to the predictions. Some tree
learning algorithms also allow for inequality tests and can work
on noisy and incomplete datasets. Decision tree learning can be
successfully applied when the samples can be described via a fixed
number of parameters and when the output forms a discretised set.
For instance, Figure \ref{decisionTreeFigure} presents the learnt
concept of \emph{ShuttleLaunch} which suggests whether it is
suitable to proceed with a scheduled launching after taking into
account system checks and forecasted weather.

Like other supervised learning algorithms, a set of training
examples with known classification is initially processed to infer
a decision tree.  Hopefully, this would be a good description of
the classification procedure and will eventually be used to
distinguish the class of other unseen samples. The learnt class
description can be understood by humans and new knowledge about
galaxy morphology classification in particular, can be obtained.

The approach adopted by most algorithms is very much the same.
Moving in a top down manner, a greedy search that goes through the
entire space of possible trees is initiated to try and find the
minimum structure that correctly represents the examples. The
constructed tree is then used as a rule set for predicting the
class category of an unknown sample from the same set of
attributes.

\subsection{C4.5}
Following work done by Hunt in the late 1950s and early 1960s,
Ross Quinlan continued to improve on the developed techniques and
released the Iterative Dichottomizer 3 (ID3) and the improved C4.5
decision tree learners \citep{kohavi90}. Even though the C5
algorithm is commercially available, the freely available C4.5
algorithm will be brought forward.

Here trees are built by recursively searching through and
splitting the provided training set.  If all samples in the set
belong to the same class, the tree is taken to be made of just a
leaf node. Otherwise, the values of the parameters are tested to
determine a non trivial partition that separates the samples into
the corresponding classes. In C4.5, the selected splitting
criterion is the one that maximizes the information gain and the
gain ratio.

Let $RF(C_{j}, S)$ be the relative frequency of samples in set $S$
that belongs to class $C_{j}$. The information that identifies the
class of a sample in set $S$ is:
\begin{displaymath}
I(S) = -\sum_{j=1}^xRF(C_{j}, S) log (RF(C_{j}, S))
\end{displaymath}
After applying a test $T$ that separates set $S$ in $S_{1}, S_{2},
..., S_{n}$, the information gained is:
\begin{displaymath}
G(S, T) = I(S) - \sum_{i=1}^{t} \frac{|S_{i}|}{|S|}I(S_{i})
\end{displaymath}
The test that maximises $G(S, T)$ is selected at the respective
node. The main problem with this approach is that it favours tests
having a large number of outcomes such as those producing a lot of
subsets each with few samples. Hence the gain ratio that also
takes the potential information from the partition itself is
introduced:
\begin{displaymath}
P(S, T) = - \sum_{i=1}^{t}
\frac{|S_{i}|}{|S|}log\left(\frac{|S_{i}}{|S|}\right)
\end{displaymath}

If all samples are classified correctly, the tree may be
overfitting the data and will fail when attempting to classify
more general, unseen samples. Normally this is prevented by
restricting some examples from being considered when building the
tree or by pruning some of the branches after the tree is
inferred. C4.5 adopts the latter strategy and remove some branches
in a single bottom up pass.

One of the main advantages of the C4.5 algorithm is that it is
capable of dealing with real, non-nominal attributes and so
renders itself compatible with continuous parameters. It can also
handle missing attribute data.

\subsection{CART}
The Classification and Regression Tree (CART) scheme was developed
by Friedman and Breiman \citep{kohavi90}. Although a divide and
conquer search strategy similar to that of C4.5 is used, the
resulting tree structure, the splitting criteria, the pruning
method as well as the way missing values are handled, are
redefined.

CART only allows for binary trees to be created. While this may
simplify splitting and optimally partitions categorical
attributes, there may be no good binary split for a parameter and
inferior trees might be inferred. However, for multi-class
problems, twoing may be used. This involves separating all samples
in two mutually exclusive super-classes at each node and then
apply the splitting criteria for a two class problem.

As a splitting criterion, CART uses the Gini diversity index. Let
$RF(C_{j}, S)$ again be the relative frequency of samples in set
$S$ that belongs to class $C_{j}$, then the Gini index is defined
as:
\begin{displaymath}
I_{gini}(S) = 1 - \sum_{j=1}^xRF(C_{j}, S)^{2}
\end{displaymath}
and the information gain due to a particulat test $T$ can be
computed from:
\begin{displaymath}
G(S, T) = I_{gini}(S) - \sum_{i=1}^{t}
\frac{|S_{i}|}{|S|}I_{gini}(S_{i})
\end{displaymath}
As with the C4.5 algorithm, the split that maximises $G(S, T)$, is
selected. If all samples in a given node have the same parameter
value, then the samples are perfectly homogenous and there is no
impurity.

The CART algorithm also prunes the tree and use cross validation
methods that may require more computation time. However, this will
render shorter trees than those obtained from C4.5. Samples with
missing data may also be processed.

\subsection{Random Forests}
\cite{breiman}, the pioneers of random forests, suggest using a
classifier in which a number of decision trees are built. When
processing a particular sample, the output by each of the
individual trees is considered and the resulting mode is taken as
the final classification. Each tree is grown from a different
subset of examples allowing for an unseen (out of bag) set of
samples to be used for evaluation. Attributes for each node are
chosen randomly and the one which produces the highest level of
learning is selected. It is shown that the overall accuracy is
increased when the trees are less correlated. Having each of the
individual trees with a low error rate, is also desirable.

Apart from producing a highly accurate classifier, such a scheme
can also handle a very large amount of samples and input
variables. A proximity matrix which shows how samples are related,
is also generated. This is useful since such relations may be very
difficult to be detect just by inspection. With this strategy,
good results may still be obtained even when a large portion of
the data is missing or when the number of examples in each
category is biased.

\section{Fuzzy Inference Systems}
\label{fuzzyInferenceSystemsSection}

Fuzzy logic accommodates soft computing by allowing for an
imprecise representation of the real world. In crisp logic a clear
boundary is considered to separate the various classes and each
element is categorised into one group such that samples in sets
$A$ and $not A$ represents the entire dataset. Fuzzy logic extends
on this by giving all sample a degree of membership in each set
hence also caters for situations in which simple boolean logic is
not enough. If classically set membership was denoted by 0 (false)
or 1 (true), now we can also have $0.25$ or $0.75$. In fuzzy
logic, the truth of any statement becomes a matter of degree.

The mathematical function that maps each input to the
corresponding membership value between $0$ and $1$ is known as the
membership function. Although this can be arbitrary, such function
is normally chosen with computation efficiency and simplicity kept
in mind.  Various common membership functions include the
triangular function, the trapezoidal function, the gaussian
function and the bell function. The latter are the most popular
and although they are smooth, concise and can attain non-zero
values anywhere, they fail in specifying asymmetric membership
functions. Such limitation is elevated through the use of sigmoid
functions that can either open left or right.

For an inference system, $if-then$ rules that deal with fuzzy
consequents and fuzzy antecedents are defined. An aggraded fuzzy
set is then outputted after these conditional rules are compared
and combined by standard logical operators equivalents. Since the
degree of membership can now attain any value between $0$ and $1$,
the $AND$ and $OR$ operators are replaced by the $max$ and $min$
functions respectively. The resulting output is then defuzzified
to obtain one output value.

Fuzzy inference systems are easily understood and can even be
applied when dealing with imprecise data. Like decision tree
classifiers, they provide a penetrative model that experts can
analyze and even add other information to it. Such inference
approaches have already been successfully applied in a number of
applications that range from integration in consumer produces to
industrial process control, medical instrumentation and decision
support systems.

\section{Input Parameters}
\label{inputParametersSection}

In all machine learning algorithms, the set of input parameters
strongly determine the overall accuracy of the classifier.
Ideally, a minimum number of attributes that can differentiate
between the three galaxy morphology classes are required. For this
work, photometric and spectra values downloaded from the SDSS
\texttt{PhotoObjAll} and \texttt{SpecLineAll} tables, were used.
Data for which classification information is available in the
Galaxy Zoo catalogue were downloaded and used to test the various
machine learning algorithms used.

\subsection{Photometric Attributes}

In this study, the set of 13 parameters as taken by
\cite{banerji09} which are based on colour, profile fitting and
adaptive moments were used. However, we did not limit the
evaluation to the \emph{i} band but also aimed at testing whether
the values derived from the \emph{r} band give equal or better
classification accuracies. The input parameters used are presented
in Table \ref{parametersTable}.

\begin{table}
\caption{Set of input parameters that are band independent, from
the \emph{i} band ($\approx700nm - 1400nm$) and from the \emph{r}
band ($\approx700nm$)}\label{parametersTable}
\begin{tabular}{@{}ll}
\hline
Name & Description\\
\hline
\texttt{dered\_g} - \texttt{dered\_r} & deredded (g - r) colour\\
\texttt{dered\_r} - \texttt{dered\_i} & deredded (r - i) colour\\
\hline
\texttt{deVAB\_i} & DeVaucouleurs fit axis ratio\\
\texttt{expAB\_i} & Exponential fit axis ratio\\
\texttt{lnLExp\_i} & Exponential disk fit log likelihood\\
\texttt{lnLDeV\_i} & DeVaucouleurs fit log likelihood\\
\texttt{lnLStar\_i} & Star log likelihood\\
\texttt{petroR90\_i} / \texttt{petroR50\_i} & Concentration \\
\texttt{mRrCc\_i} & Adaptive (+) shape measure\\
\texttt{texture\_i} & Texture parameter\\
\texttt{mE1\_i} & Adaptive E1 shape measure\\
\texttt{mE2\_i} & Adaptive E2 shape measure\\
\texttt{mCr4\_i} & Adaptive fourth moment\\
\hline
\texttt{deVAB\_r} & DeVaucouleurs fit axis ratio\\
\texttt{expAB\_r} & Exponential fit axis ratio\\
\texttt{lnLExp\_r} & Exponential disk fit log likelihood\\
\texttt{lnLDeV\_r} & DeVaucouleurs fit log likelihood\\
\texttt{lnLStar\_r} & Star log likelihood\\
\texttt{petroR90\_r} / \texttt{petroR50\_r} & Concentration \\
\texttt{mRrCc\_r} & Adaptive (+) shape measure\\
\texttt{texture\_r} & Texture parameter\\
\texttt{mE1\_r} & Adaptive E1 shape measure\\
\texttt{mE2\_r} & Adaptive E2 shape measure\\
\texttt{mCr4\_r} & Adaptive fourth moment\\
\end{tabular}
\end{table}

The DeVaucouleurs law provides a measure of how the surface
brightness of an elliptical galaxy varies with apparent distance
from the centre. This should provide a good element of
discrimination between spiral and elliptical profiles. The
\texttt{lnLStar} parameter also helps to separate galaxy from star
objects. The concentration parameter is given by the ratios of
radii containing 90\% and 50\% of the Petrosian flux in a given
band. The \texttt{texture} parameter compares the range of
fluctuations in the surface brightness of the object to the full
dynamic range of the surface brightness. It is expected that this
is negligible for smooth profiles but becomes significant in high
variance regions such as spiral arms.

The other parameters used are based on the object's shape.
Particularly, the adaptive moments derived from the SDSS
photometric pipeline are second moments of the object intensity,
measured using a particular scheme designed to have an optimal
signal to noise ratio. These moments are calculated by using a
radial weight function that adopts to the shape and size of the
object. Although theoretically there exists an optimal radial
shape for the wight function related to the light profile of the
object, a Gaussian with size matched to that of the object is used
\citep{bernstein02}.

The sum of the second moments in the CCD row and column direction
(\texttt{mRrCc}) is calculated by:
\begin
{displaymath} mRrCc = <c^{2}> + <r^{2}>
\end{displaymath}
where $c$ and $r$ correspond to the columns and rows of the sensor
respectively and the second moments are defined as
\begin{displaymath}
<c^{2}>=\frac{\sum[I(r, c)w(r, c)c^{2}]}{\sum[I(r, c)w(r, c)]}
\end{displaymath}
$I$ is the intensity of the object and $w$ is the weighting
function. The ellipticity/polarisation components are defined by:
\begin{displaymath}m_{e1} = <c^{2}> - \frac{<r^{2}>}{MRrCc}\end{displaymath}
\begin{displaymath}m_{e2} = 2\frac{<(c)(r)>}{MRrCc}\end{displaymath}
A fourth order moment is also defined as:
\begin{displaymath}m_{cr4} = \frac{<(c^{2} + r^{2})^{2}>}{\sigma^{4}}\end{displaymath}
In this case, $\sigma$ is the weight of the Gaussian function
applied.

\subsection{Spectra Attributes}
\label{spectraAttributesSubSection}

To try and differentiate between elliptical, spiral and other
morphologically shaped galaxies, this study also makes use of
strong emission lines captured in galactic spectra. Significant
lines of oxygen and hydrogen around the 5000\AA and 7000\AA marks
are expected for spiral galaxies. Since these often have star
forming regions in the arms, the presence of sulphur, nitrogen and
helium originating from ionized gas clouds, is also expected.
Elliptical galaxies on the other hand are believed to have no star
forming activity and can therefore be identified from continuous
and average spectra.

Part of the SDSS pipeline is responsible to detect and store all
strong emission lines present in the captured spectra. This is
achieved through wavelet filters. An attempt to match all peaks
with one of the candidate emission lines defined in a list, is
made. Each line is then fitted with a single Gaussian by the
SLATEC common mathematical library routine SNLS1E
\citep{spectraSdss}. The height and the dispersion of the fitted
Gaussian, the resulting Chi-squared error and other derived
parameters are stored in the corresponding database. Table
\ref{spectraTable} presents the wavelengths of lines considered
for this study. Lines storing missing dummy values for more than
5\% of the dataset, were ignored. Although still randomly
selected, samples for the training and testing sets were biased
towards entries with a lower chi-squared error. This allowed for
height values of better fitted Gaussians to be considered.

\begin{table}
\centering
\caption{Wavelengths of spectra lines}
\label{spectraTable}
\begin{tabular}{@{}lllll}
\hline
Wave & Label && Wave & Label\\
\hline
3727.09   & OII 3727     & & 4960.30   & OIII 4960\\
3729.88   & OII 3730     & & 5008.24   & OIII 5008\\
3798.98   & Hh 3799      & & 5176.70   & Mg 5177\\
3836.47   & Oy 3836      & & 5895.60   & Na 5896\\
3889.00   & HeI 3889     & & 6302.05   & OI 6302\\
3934.78   & K 3935       & & 6365.54   & OI 6366\\
4072.30   & SII 4072     & & 6549.86   & NII 6550\\
4102.89   & Hd 4103      & & 6564.61   & Ha 6565\\
4305.61   & G 4306       & & 6585.27   & NII 6585\\
4341.68   & Hg 4342      & & 6718.29   & SII 6718\\
4364.44   & OIII 4364    & & 6732.67   & SII 6733\\
4862.68   & Hb 4863      & &           &\\
\end{tabular}
\end{table}

\section{Results}
\label{resultsSection}

\begin{figure}
\centering
\includegraphics[width=84mm]{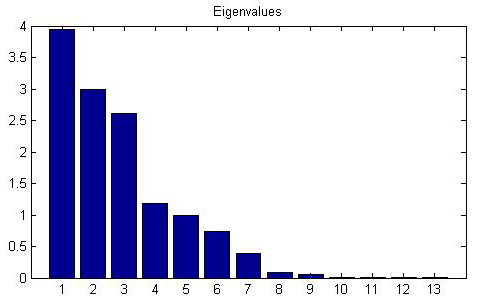}
\caption {Eigenvalues from the 13 \emph{i} band parameters}
\label{eigenvaluesI}
\end{figure}

\begin{figure}
\centering
\includegraphics[width=84mm]{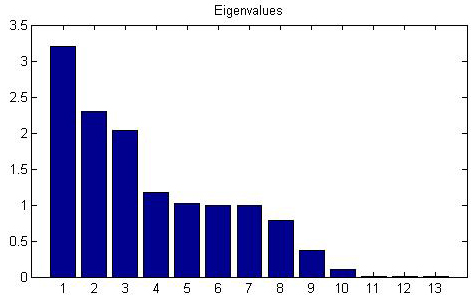}
\caption {Eigenvalues from the 13 \emph{r} band parameters}
\label{eigenvaluesR}
\end{figure}

\begin{figure}
\centering
\includegraphics[width=84mm]{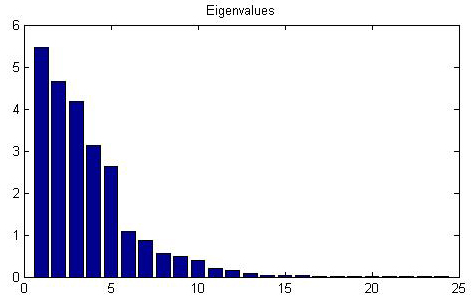}
\caption {Eigenvalues from the 24 \emph{i} and \emph{r} band
parameters} \label{eigenvaluesIR}
\end{figure}

Initially, the 13 photometric parameters derived from the \emph{i}
band were standardised and independent component analysis was
performed to determine the most significant components. As can be
seen from the resulting eigenvalues shown in Figure
\ref{eigenvaluesI}, all of the independent components attain a
non-zero value. This implies that all attributes are important for
galaxy classification and dimension reduction is unnecessary.
Figure \ref{eigenvaluesR} and Figure \ref{eigenvaluesIR} show the
eigenvalues obtained when the same analysis was repeated on
\emph{r} band data as well as on the 24 attributes obtained when
the \emph{i} and \emph{r} bands parameters were combined.

Once the significance of the selected parameters was confirmed,
various machine learning algorithms were tested. In particular,
the CART algorithm, the C4.5 algorithm with confidence values of
0.25 and 0.1 and the Random Forest algorithm with 10 and 50 trees
were considered. For all test cases, a ten-fold cross validation
strategy was used. The compiled 75,000 sample set (Set 3) was
divided into 10 complimentary subsets and the learning algorithm
was executed for 10 times. In each run, one of the ten subsets was
used for testing and the other nine subsets were put together to
form the training set. The presented results are the computed
averages across all ten trials. By this approach, every sample is
part of the test set at least once.

\subsection{\emph{i} Band Photometric Parameters}
\label{nearInfraredWavebandPhotometricParametersSubSection}

\begin{figure}
\centering
\includegraphics[width=84mm]{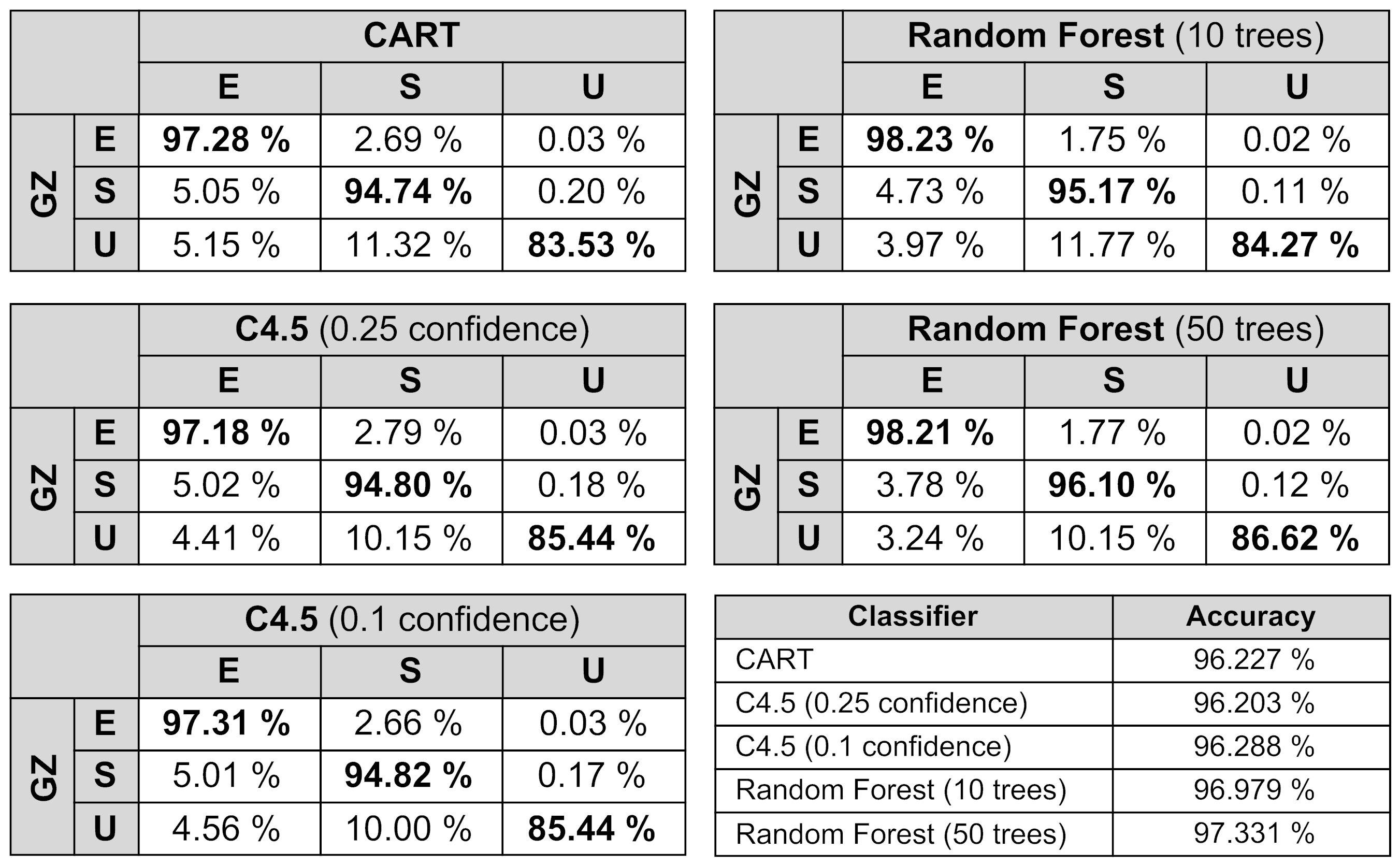}
\caption {Decision tree confusion matrices for the \emph{i} band
input parameters} \label{resultsGoldSubI}
\end{figure}

The resulting confusion matrices when considering the 13 \emph{i}
band parameters are shown in Figure \ref{resultsGoldSubI}. These
tables correlate the actual morphological classes with those
outputted by the classifier. For instance, in the first row, the
percentages of elliptical galaxies that were classified as
elliptical (E), spiral (S) and unknown (U) are shown. Decision
trees output a single morphology type for every input, therefore
the percentages in each row add up to 100\%. This corresponds to
all of the input samples of a particular class. The global
accuracy percentage was then calculated by comparing the total
number of correctly classified samples with the total number of
inputted tests.

In all decision tree algorithms tested, the global accuracy is
always above 96.2\% with the highest being 97.33\% achieved by the
50 tree random forest technique. All confusion matrices result to
have the highest values in the diagonal. This indicates that the
majority of samples were classified correctly. The random forest
algorithm with 50 trees did prove to be the most accurate and did
manage to correctly classify 98.21\% of all ellipticals, 96.10\%
of all spirals and 86.62\% of all unknown objects. The slightly
less than optimal classification percentages for unknown objects
can be due to a number of factors. First of all, the number of
training samples with unknown morphology might have not been
enough for the algorithm to learn how to identify such samples and
secondly, objects that mislead humans might actually have very
similar properties to spiral or elliptical galaxies and are
ultimately also classified correctly by the algorithm.

\begin{figure}
\centering
\includegraphics[width=50mm]{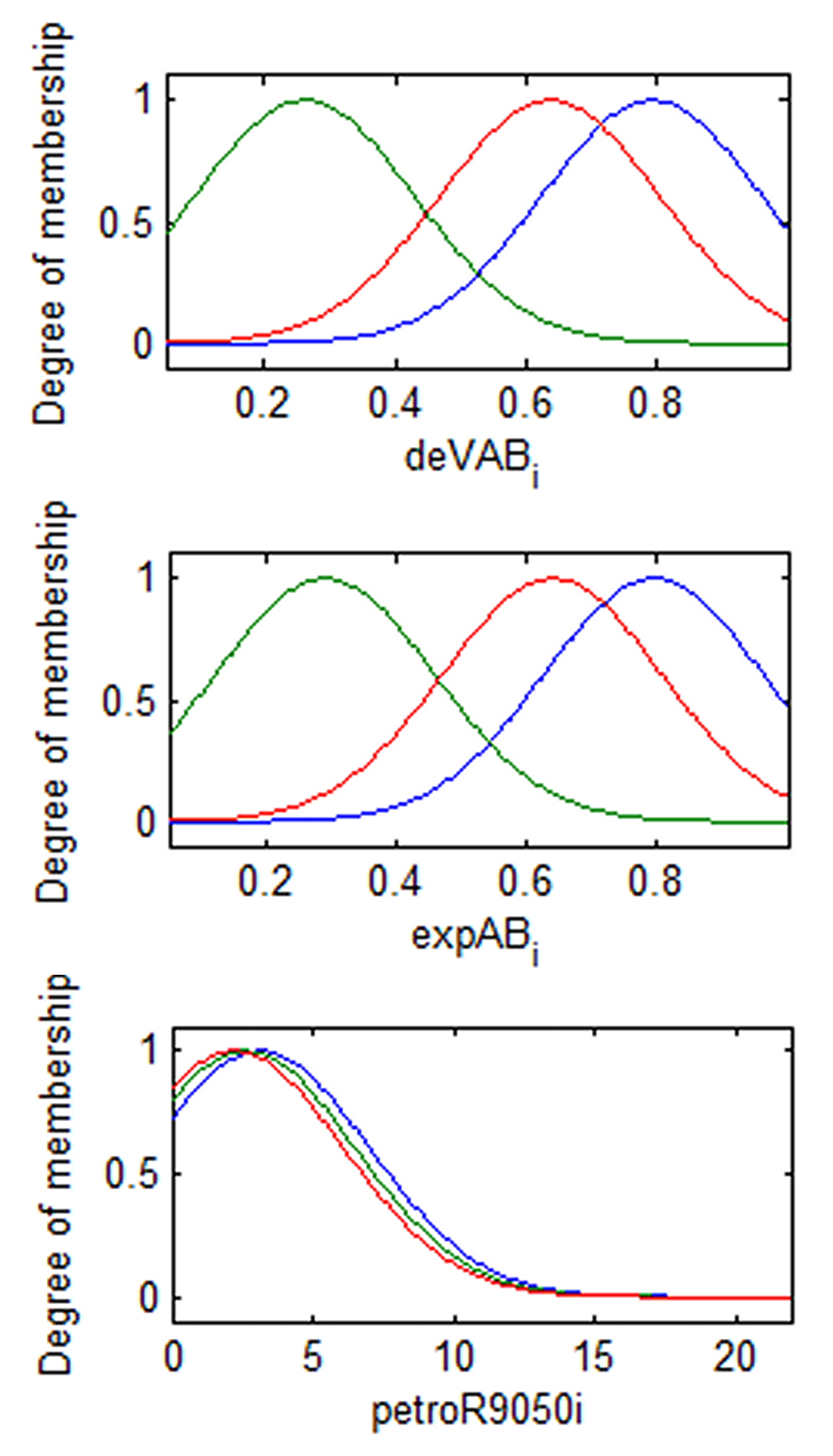}
\caption {Fuzzy inference system membership functions}
\label{fisMembershipFunctions}
\end{figure}

\begin{figure}
\centering
\includegraphics[width=43mm]{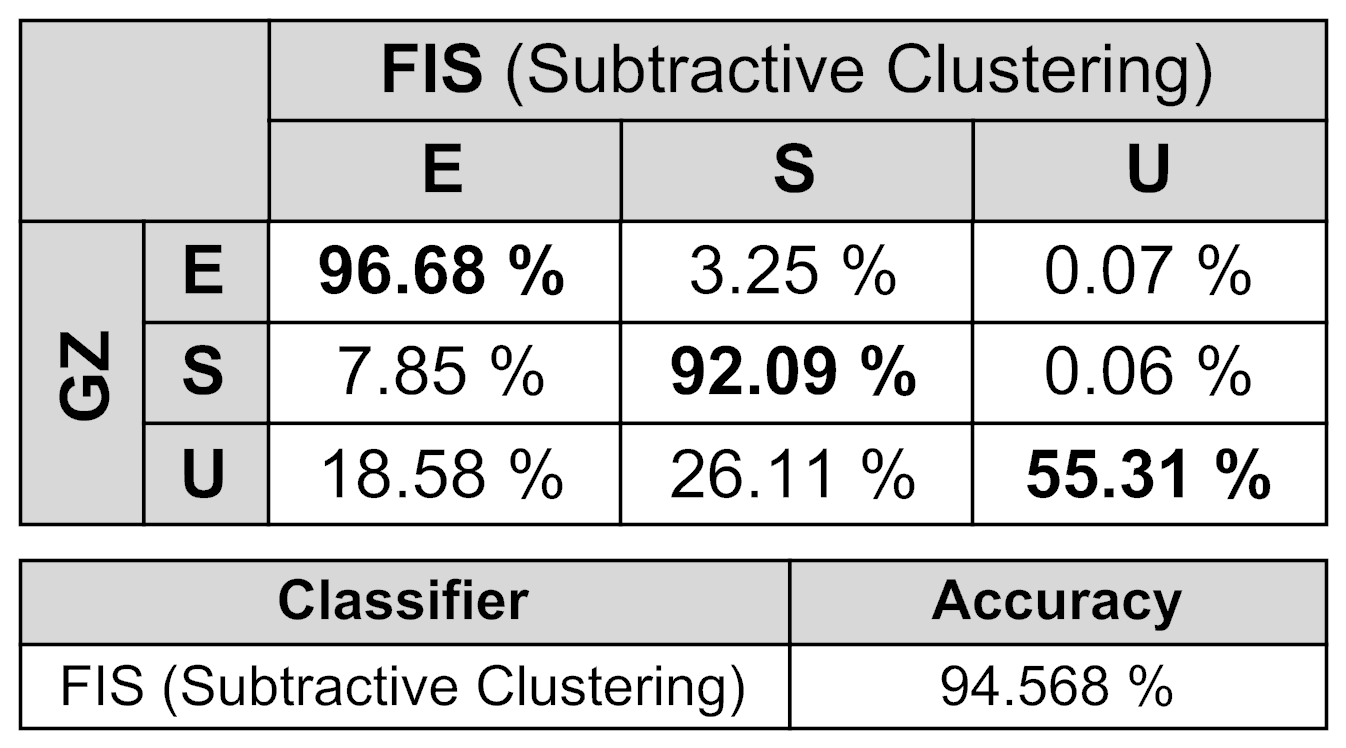}
\caption {Fuzzy inference system confusion matrix for \emph{i}
band input parameters} \label{resultsGoldSubFisI}
\end{figure}

The membership functions derived by the fuzzy inference system for
the DeVaucouleurs fit axis ratio, exponential fit axis ratio and
concentration parameters in the \emph{i} band, are shown in Figure
\ref{fisMembershipFunctions}. For such a model, subtracting
clustering was used. The results obtained after testing are
presented in Figure \ref{resultsGoldSubFisI}. Clearly, the
developed model is capable of describing elliptical and spiral
galaxies but suffers to accurately detect galaxies tagged to have
an unknown type.

\subsection{\emph{r} Band Photometric Parameters}
\label{redWavebandPhotometricParametersSubSection}

\begin{figure}
\centering
\includegraphics[width=84mm]{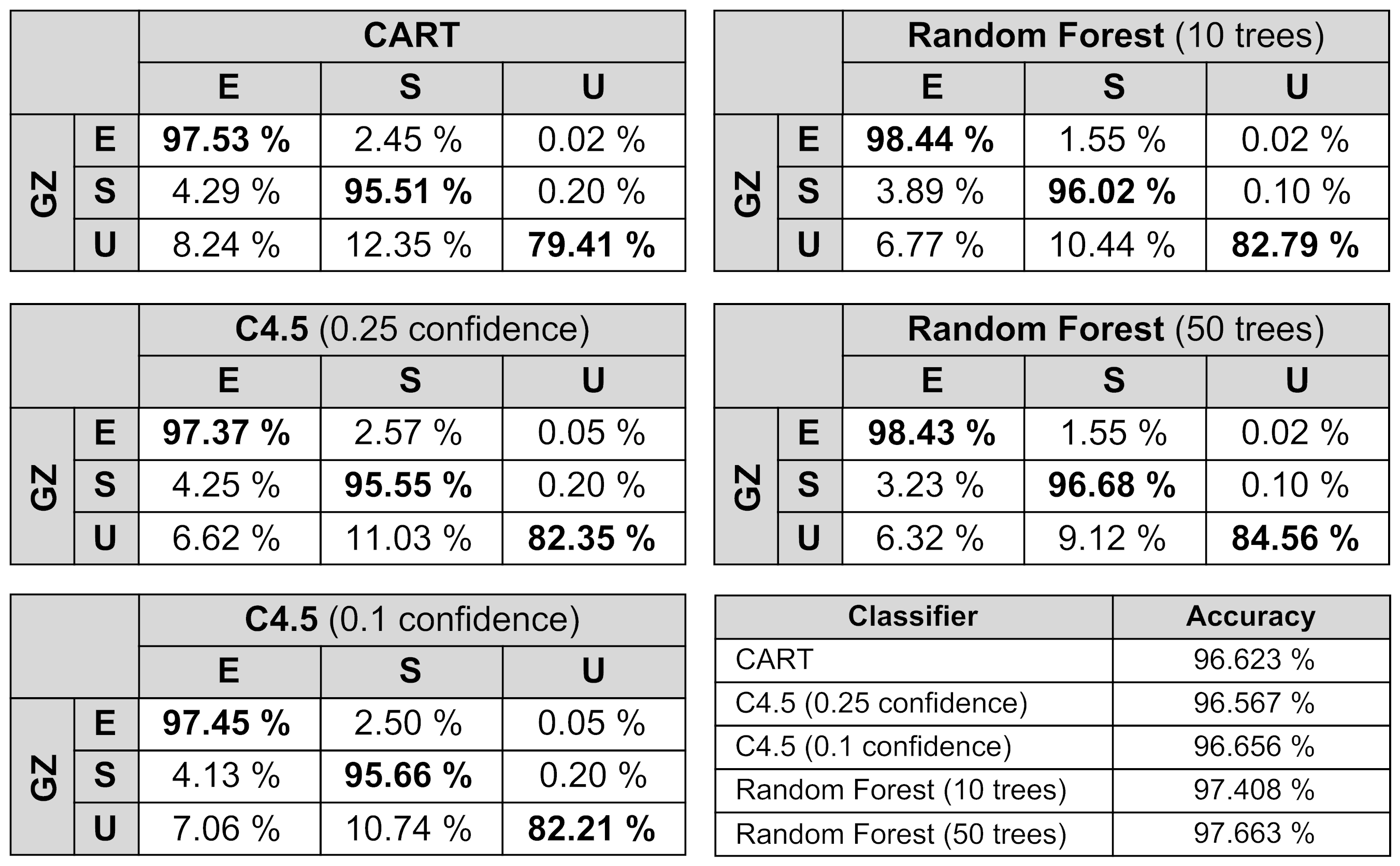}
\caption {Decision tree confusion matrices for \emph{r} band input
parameters} \label{resultsGoldSubR}
\end{figure}

The CART, C4.5 and Random Forest decision tree algorithms were
also applied to objects in Set 3 with the input parameters
extracted from the \emph{r} band. The resulting confusion matrices
are presented in Figure \ref{resultsGoldSubR}. When compared to
the results obtained from the \emph{i} band data, a gain in the
general accuracies of all algorithms can be noted. Using the
\emph{r} band information seems to help in distinguishing between
ellipticals and spirals. However, the same cannot be said for the
unclassified class since accuracies for this morphology class
depreciated from an overall average of about 81\% to that of about
75\%.

\subsection{\emph{i} and \emph{r} Bands Photometric Parameters}
\label{nearInfraredAndRedWavebandPhotometricParametersSubSection}

\begin{figure}
\centering
\includegraphics[width=84mm]{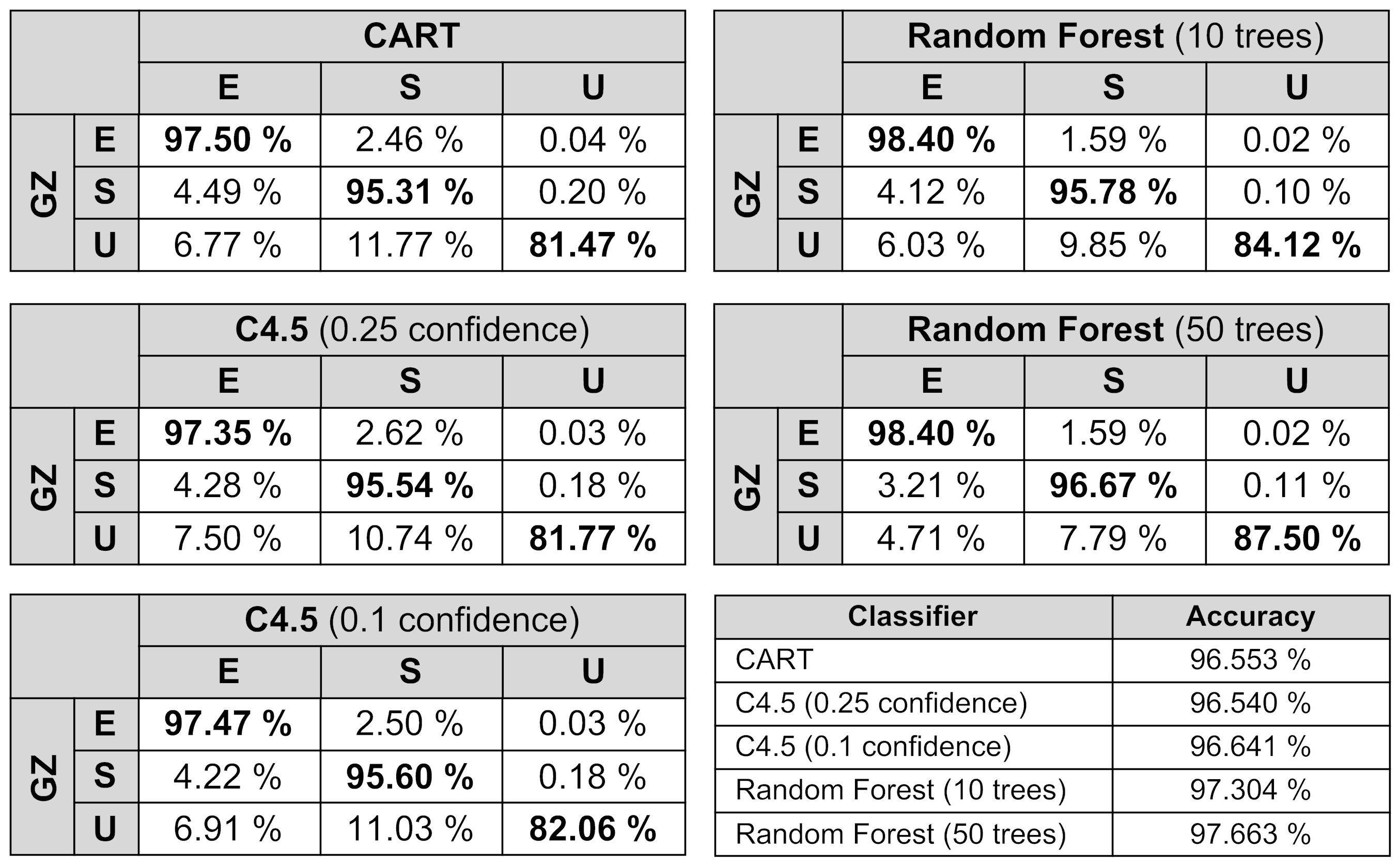}
\caption {Decision tree confusion matrices for \emph{i} and
\emph{r} bands input parameters} \label{resultsGoldSubIR}
\end{figure}

Following the tests described in Section
\ref{nearInfraredWavebandPhotometricParametersSubSection} and
Section \ref{redWavebandPhotometricParametersSubSection} above,
models built from the \emph{i} and \emph{r} bands parameters were
tested. A 24 attribute dataset was constructed for objects defined
in Set 3 and the same decision tree algorithms were applied to
obtain corresponding classifiers. The results are presented in
Figure \ref{resultsGoldSubIR}. Although an improvement in accuracy
is registered by the 50 tree Random Forest algorithm, this is only
by a very small percentage.

\subsection{Spectra Parameters}

\begin{figure}
\centering
\includegraphics[width=84mm]{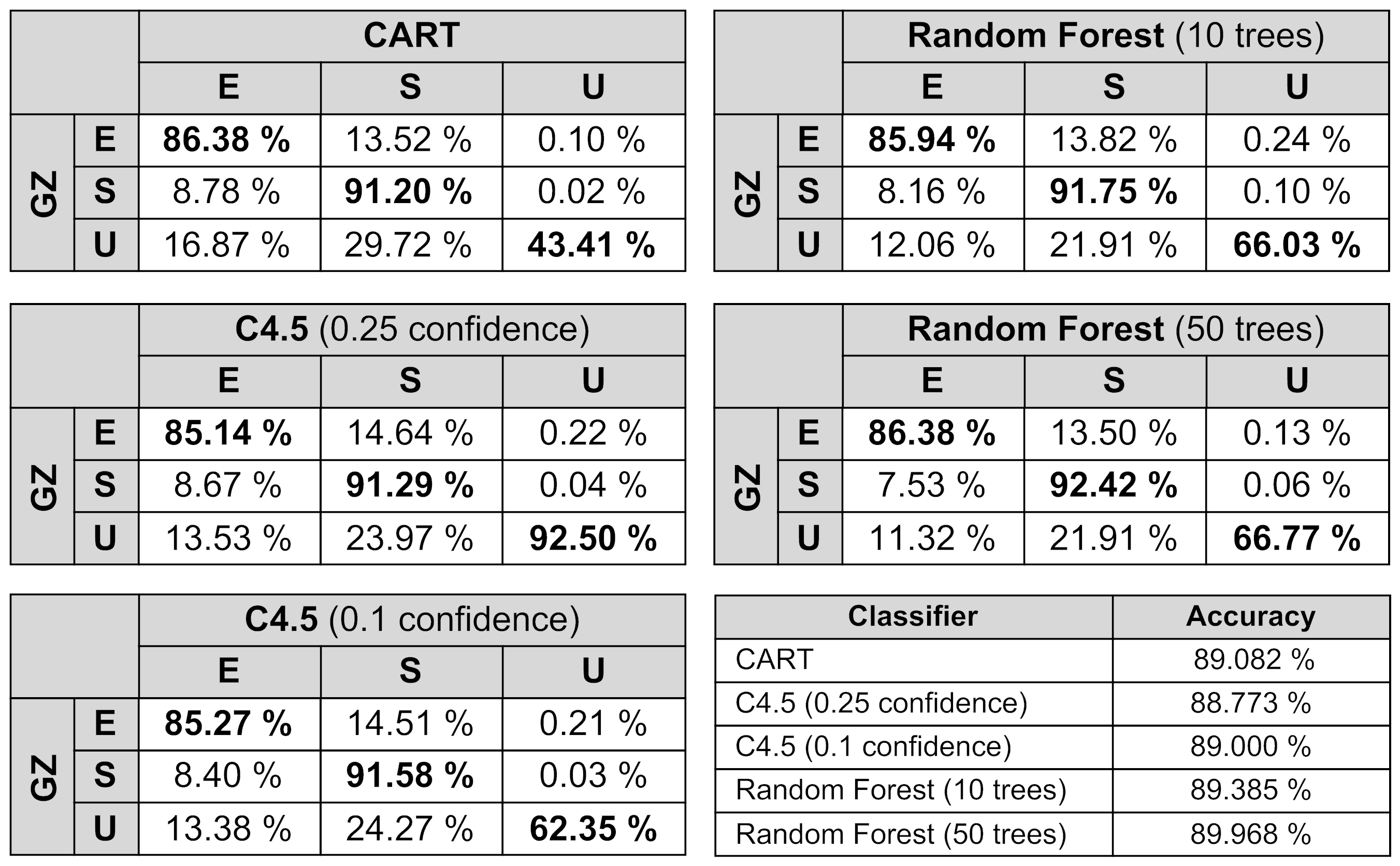}
\caption {Decision tree confusion matrices for spectra input
parameters} \label{resultsGoldSubSpectra}
\end{figure}

A similar methodology was adopted to test classification
accuracies of models developed from spectra data. All wave line
entries for objects in Set 3 were initially downloaded from the
SDSS database. As described in Section
\ref{spectraAttributesSubSection}, the wavelengths for which more
than 95\% of the data was available, were considered. This allowed
for a 24 attribute feature space and the achieved results are
presented in Figure \ref{resultsGoldSubSpectra}. Although still
reasonably accurate, the general classification capability was
found to be less than that obtained when photometric parameters
were used. This could be due to the fact that peak spectral lines
which are significant to detect spiral galaxies, are not always
present.

\section{Conclusion}
\label{conclusionSection}

\begin{figure}
\centering
\includegraphics[width=70mm]{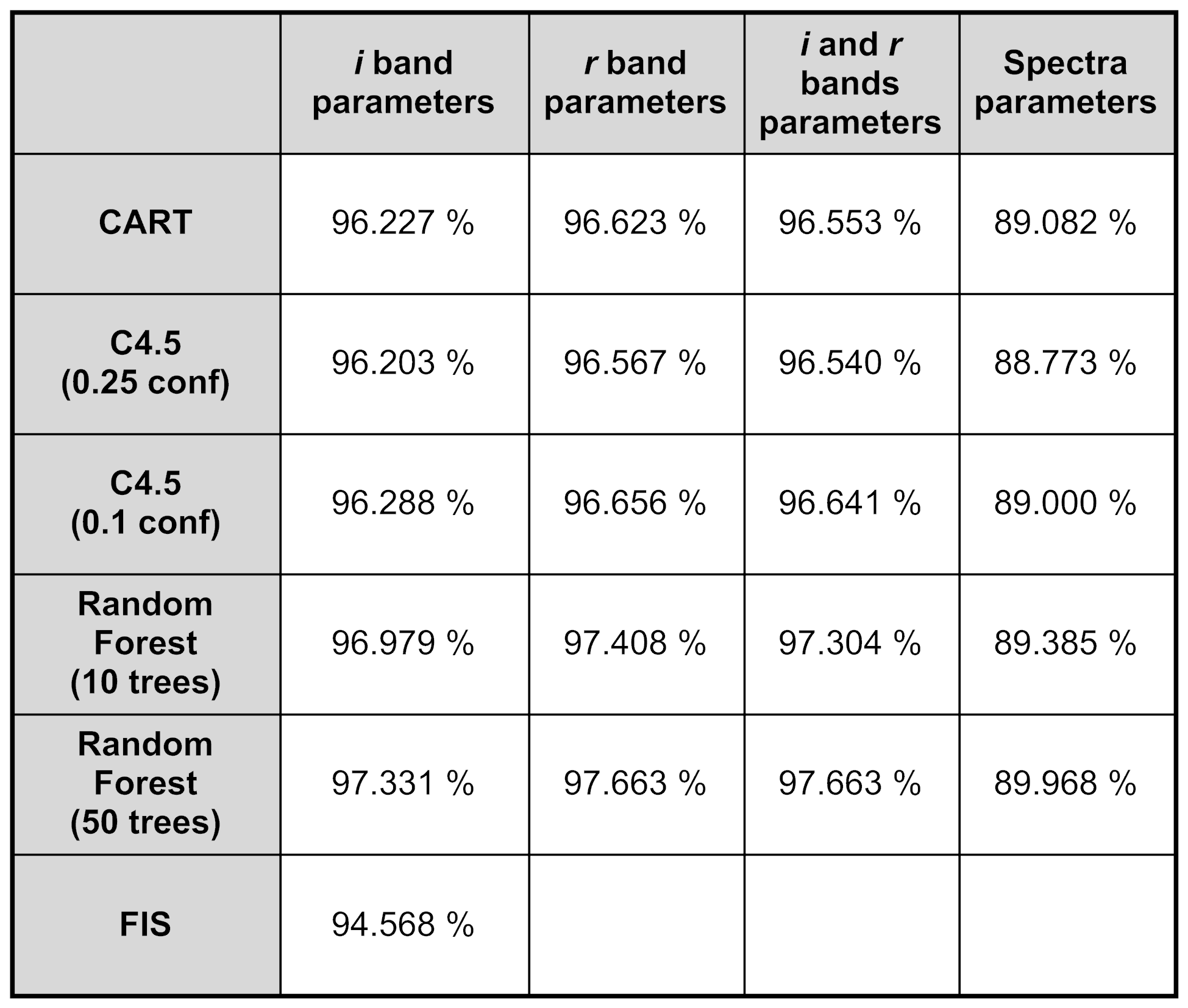}
\caption {Data, algorithms and results}
\label{resultsGeneralSummary}
\end{figure}

In this study, accuracies for different galaxy morphology
classification models developed through various machine learning
techniques were obtained and analyzed. Results from the CART, the
C4.5 and the Random Forest decision tree algorithms as well as the
output from Fuzzy Inference Systems, were compared. The advantages
gained by performing computations on different photometric
parameters and on spectra attributes, were also investigated and
put forward. Figure \ref{resultsGeneralSummary} serves as a good
summary of which data and algorithms were used as well as the
overall accuracies obtain. In all cases, the Random Forest gave
the highest percentages especially when 50 trees were used.

All of the tested algorithms took only a few minutes to run on a
normal personal computer. Although the presented results are for
Set 3, experiments on Set 2 that stored more samples were also
carried out. Accuracy percentages very close to the ones
published, were obtained.

In most cases, when processing photometric parameters, the
adaptive shape measure (\texttt{mRrCc}) parameter was chosen as
the root of the tree. First level nodes included the concentration
(\texttt{petroR90}/\texttt{petroR50}) and the
\texttt{dered\_g}-\texttt{dered\_r} parameters. For spectra data,
the Ha wave line was determined to provide the highest information
gain while the Hb and the K lines were chosen as first level
nodes.

\begin{figure}
\centering
\includegraphics[width=84mm]{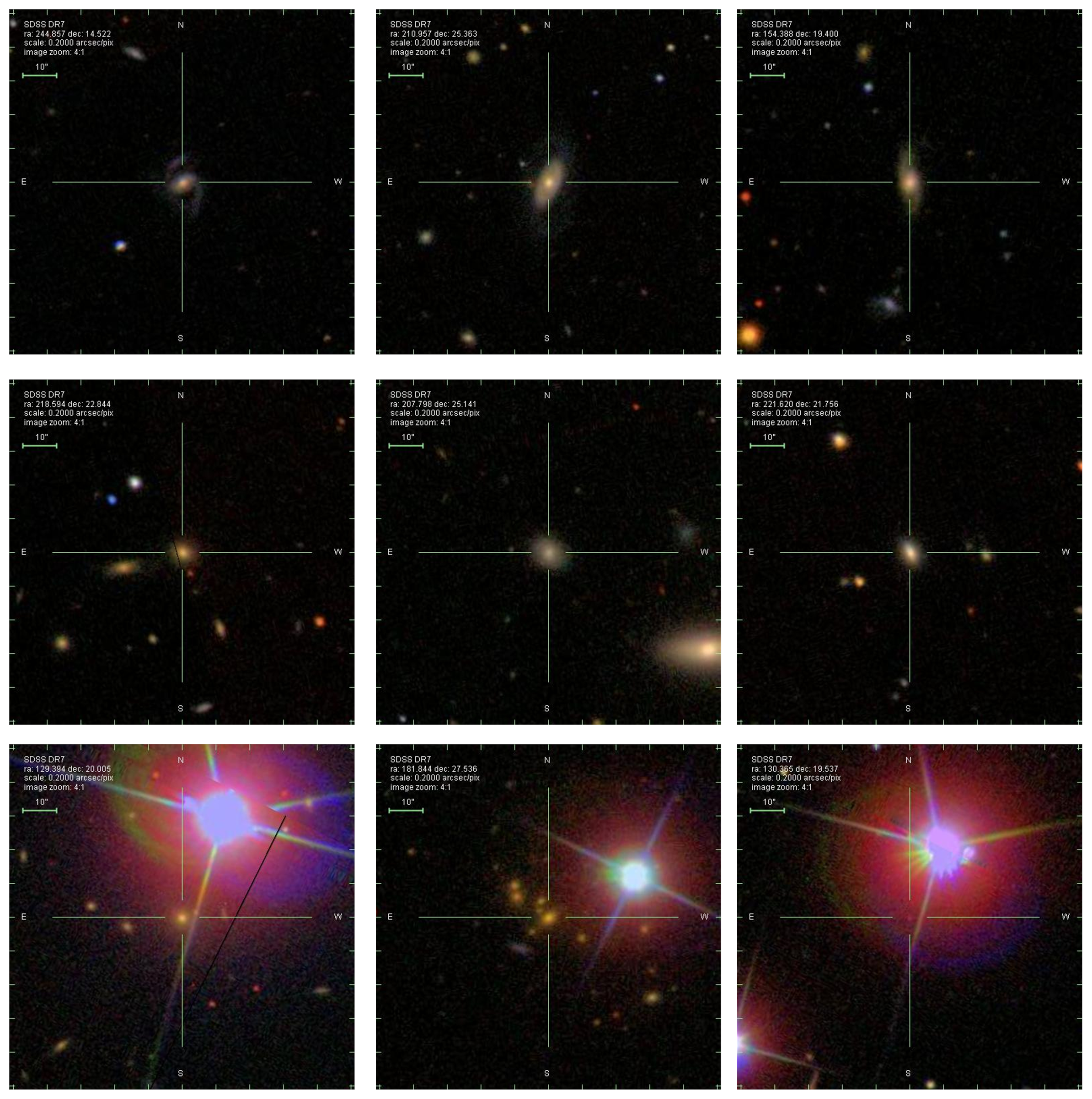}
\caption {Samples of spiral (top), elliptical (middle) and unknown
(bottom) galaxies that were incorrectly classified by the fuzzy
inference system} \label{incorrectClassification}
\end{figure}

Figure \ref{incorrectClassification} shows samples of incorrectly
classified galaxies by the fuzzy inference system. Although this
is not the most accurate technique described, the incorrectly
classified spiral and elliptical samples are very faint in
magnitude. Moreover, all incorrectly classified unknown objects
have bright sources in the vicinity and this could have had an
effect on the calculated parameters by the SDSS photometric
pipeline.

\section{Acknowledgements}
The GalaxyZoo data was supplied by Dr Steven Bamford on behalf of
the Galaxy Zoo team. The authors would like to thank him for his
comments and suggestions that helped to improve this paper.

Funding for the SDSS and SDSS-II has been provided by the Alfred
P. Sloan Foundation, the Participating Institutions, the National
Science Foundation, the U.S. Department of Energy, the National
Aeronautics and Space Administration, the Japanese Monbukagakusho,
the Max Planck Society, and the Higher Education Funding Council
for England. The SDSS Web Site is http://www.sdss.org/.

The SDSS is managed by the Astrophysical Research Consortium for
the Participating Institutions. The Participating Institutions are
the American Museum of Natural History, Astrophysical Institute
Potsdam, University of Basel, University of Cambridge, Case
Western Reserve University, University of Chicago, Drexel
University, Fermilab, the Institute for Advanced Study, the Japan
Participation Group, Johns Hopkins University, the Joint Institute
for Nuclear Astrophysics, the Kavli Institute for Particle
Astrophysics and Cosmology, the Korean Scientist Group, the
Chinese Academy of Sciences (LAMOST), Los Alamos National
Laboratory, the Max-Planck-Institute for Astronomy (MPIA), the
Max-Planck-Institute for Astrophysics (MPA), New Mexico State
University, Ohio State University, University of Pittsburgh,
University of Portsmouth, Princeton University, the United States
Naval Observatory, and the University of Washington.

\label{lastpage}

\bibliographystyle{plainnat}
\bibliography{bibfile}

\end{document}